\documentstyle[twoside,epsfig,my]{article} 
\long\def\prdfig#1#2#3
{
\begin{figure}
   \centering \leavevmode
   \epsfxsize=5.5in
   \epsfbox{#1}
   \vskip 0.3in
   \caption{\label{fig:#2} \normalsize #3}
   \hfill
\end{figure}
}

\catcode`\@=11
\long\def\@makefntext#1{
\protect\noindent \hbox to 3.2pt {\hskip-.9pt  
$^{{\eightrm\@thefnmark}}$\hfil}#1\hfill}		%CAN BE USED 

\def\@makefnmark{\hbox to 0pt{$^{\@thefnmark}$\hss}}	%ORIGINAL 
	
\def\ps@myheadings{\let\@mkboth\@gobbletwo
\def\@oddhead{\hbox{}
\rightmark\hfil\eightrm\thepage}   
\def\@oddfoot{}\def\@evenhead{\eightrm\thepage\hfil
\leftmark\hbox{}}\def\@evenfoot{}
\def\sectionmark##1{}\def\subsectionmark##1{}}

\font\eightrm=cmr8

\def\be{\begin{equation}}
\def\ee{\end{equation}}

\title{Search for CP violation in tau decays.}

\author{Y.~Maravin\address{Physics Department, Southern Methodist University,
Dallas, TX 75275-0175, USA\\E-mail: maravin@mail.physics.smu.edu}
\thanks{Talk given at the Frontiers in Contemporary Physics-II (FCP01)
Workshop,Vanderbilt University, Nashville TN, 5-10 March 2001.}}

\begin{document}
\begin{abstract}
The results of the searches for CP non-conservation in the decays of 
$\tau$ leptons are presented. No evidence of violation of CP symmetry 
is observed neither in CLEO nor in BELLE data. Interpretation of these
results is done within the framework of a model with a scalar
boson exchange. Limits on the imaginary part of the 
coupling constant $\Lambda$, parameterizing the relative contribution 
of diagrams that would lead to CP violation, are 
$-0.046 <\Im(\Lambda) < 0.022$ at $90\%$ C.L. and 
$|\Im(\Lambda)| < 1.7$ at $90\%$ C.L. in $\tau\to\pi\pi^0\nu_{\tau}$
and $\tau\to K\pi^0\nu_{\tau}$ modes, respectively.
\end{abstract}
\maketitle

%%%%%%%%%%%%%%%%%%%%%%%%%%%%%%%%%%%%%%%%%%%%%%%%%%%%%%%%%%%%%%%%%
\section{Introduction}
The violation of the combined symmetry of charge conjugation 
and parity (CP) has been of long-standing interest as a possible 
source of the matter-antimatter asymmetry~\cite{matter} 
in the Universe.
Efforts to search for the CP-violating effects have concentrated 
so far on the hadronic sector. 
CP violation in strange meson decay has been the subject of intensive  
investigation since its first observation~\cite{cpk} in 1964.
Studies of hadronic~\cite{direct} as well as 
semileptonic~\cite{kpimu1,kpimu2,bigi} kaon decays 
provide precision measurements of the CP violation parameters.
Searches for corresponding asymmetries in $B$~meson decays are
the focus of several large ongoing experiments~\cite{BABAR,BELLE}. 
Recent indications of possible neutrino oscillations~\cite{kamiokande} 
make it important to re-examine the question of CP non-conservation 
in the leptonic sector. Such violation is forbidden in the Standard Model 
but appears as a consequence of its various extensions~\cite{wudka}.
Among best-known theoretically are the multi-Higgs-doublet models 
(MHDM)~\cite{mhdm,mhdm2,mhdm3}. Models predicting lepton flavor 
violation often also predict CP violation in lepton decays~\cite{lfv1,lfv2}. 
Precision studies of muon decay parameters~\cite{mu_decays_work,mu_decays} 
show no indication for CP violation in such decay. 
\par
In this talk I discuss three searches for CP violation in tau 
lepton decays. The searches for the CP-violating asymmetry 
in single tau decays were done by the CLEO~\cite{colin} and 
the BELLE~\cite{belle_rho} collaborations. 
In the third analysis by CLEO~\cite{CLEO_recent} correlated tau decays 
are used to search for a non-zero value of CP-odd optimal observable 
indicating CP violation.

\vspace*{-1.8pt}   %optional

%%%%%%%%%%%%%%%%%%%%%%%%%%%%%%%%%%%%%%%%%%%%%%%%%%%%%%%%%%%%%%%%%%%%
\section{CP violation}\label{sec:cpv_lepton}
CP violation is the observable difference between a process
($a\to b$) and its CP conjugate ($\bar{a}\to \bar{b}$). The 
necessary condition to observe such a difference is to have a non-zero
CP violating phase $\phi$ in the Lagrangian describing the process.
\begin{figure}
\center
%\rule{2cm}{0.2mm}\hfill \rule{2cm}{0.2mm}
%\vskip 4cm
%\rule{2cm}{0.2mm}\hfill \rule{2cm}{0.2mm}
\psfig{figure=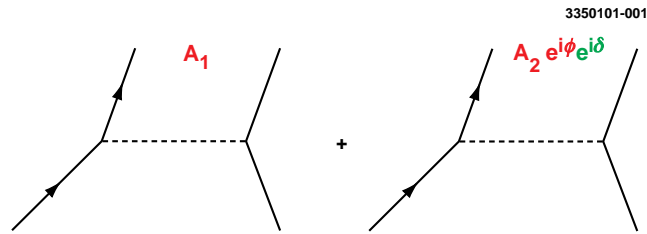,height=1.35in}
\caption{Interference 
between two amplitudes with CP-even and CP-odd relative phases 
$\delta$ and $\phi$.}
\label{fig:example}
\end{figure}
Since the absolute value of a CP-odd phase does not have a physical
meaning, CP violation can be observed only when interfered with
a CP-even phase $\delta$. Therefore, two interfering diagrams is needed
to observe CP violation (see Fig.~\ref{fig:example}). 
The probability density for such a process is given by:
$$ |{\cal A}|^2 = (A_1 + A_2 e^{i\phi} e^{i\delta})
                  (A_1 + A_2 e^{-i\phi} e^{-i\delta})$$
\begin{equation}
 = A_1^2 + A_2^2 + 2 A_1 A_2 \cos \phi \cos \delta - 
          \underline{2 A_1 A_2 \sin \phi \sin \delta}.
\end{equation}
The last, underlined, term is CP-odd since the phase $\phi$ 
changes sign under CP conjugation. If this term is not equal to zero,
physical observables between a process ($a\to b$) and its CP conjugate 
($\bar{a}\to \bar{b}$) may be different and CP is violated.
\par
$A_1$ and $A_2$ denote the amplitudes and for physical processes they
must be different from zero.  Thus CP-odd term is not equal to zero
if the factors $e^{i\phi}$ and $e^{i\delta}$ differ from zero and are 
complex.
 
%%%%%%%%%%%%%%%%%%%%%%%%%%%%%%%%%%%%%%%%%%%%%%%%%%%%%%%%%%%%%%%%%%
\section{CP violation in $\tau$ decays}
\begin{figure}
\center
\psfig{figure=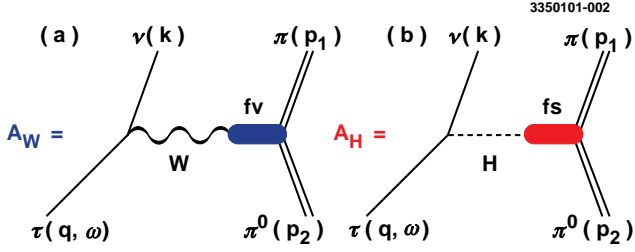,height=1.35in}
\caption{Amplitude for (a) standard W exchange, (b) scalar exchange
for $\tau\to\pi\pi^0\nu_{\tau}$ decay.}
\label{fig:Aw}
\end{figure}

A possible scenario for CP violation in $\tau$ lepton decays is
described~\cite{mhdm} by the interference of the Standard Model decay 
amplitude mediated by the W boson (amplitude $A_W$) with the amplitude
mediated by the charged Higgs boson in the multi-Higgs-doublet 
model (amplitude $A_H$). These amplitudes play the roles of $A_1$ and 
$A_2$ of the previous section (see Fig.~\ref{fig:Aw}). 
In this scenario, the charged Higgs couples to quarks and leptons with 
complex coupling constants and, thus, there can be a weak complex 
(CP-violating) phase ($\sin \phi \neq 0$). The overall Higgs
coupling to the $\pi \pi^0$ system is denoted by $\Lambda$:
\be
\Lambda = \Re(\Lambda) + i \Im(\Lambda) = |\Lambda|(\cos \phi + i \sin\phi).
\label{eq:Higgs_lambda}
\ee 
The usual choice for the CP-even phase $\delta$ is a strong 
phase~\cite{tsai2} which arises due to the QCD final state 
interactions between quarks. In the following, I consider 
only $\tau$ decays into hadronic final states and a neutrino. 
There are several possible tau decay modes that can be used to search
for CP violation. The $\tau\to K\pi \nu_{\tau}$ mode is favored with 
the largest possible CP violation effects~\cite{colin}. The corresponding 
branching fraction is, however, Cabbibo suppressed and the kaon 
identification efficiency in CLEO data is small, thus, leading to a 
small sample that can be used for experimental studies. This is not
the case for $\tau\to \pi\pi^0\nu_{\tau}$ decay due to the largest
branching fraction of tau decaying into two pion state. However, scalar
decaying into two pions implies isospin violation~\cite{tsai}. 
Therefore, the CP violating effects in this mode are 
suppressed.\footnote{The suppression factor is proportional to the mass 
difference of $u$ and $d$ quarks. It is estimated to be of the
order of $0.0035$~\cite{tsai}.} 
In the following I discuss searches for CP violation in both tau decay
modes.

%%%%%%%%%%%%%%%%%%%%%%%%%%%%%%%%%%%%%%%%%%%%%%%%%%%%%%%%%%%%%%%%%
\section{Searches for CP violation in single tau decays}
\begin{figure}
\center
\psfig{figure=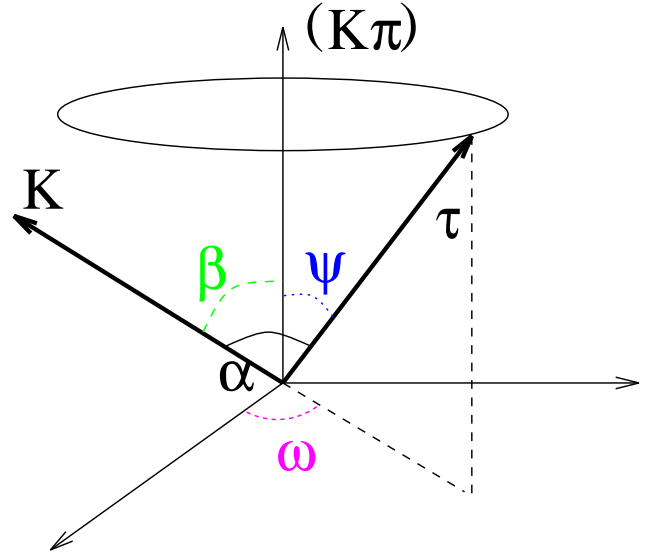,height=3in}
\caption{Definition of angles in the hadronic rest frame. ($K\pi$) 
defines the direction of the hadronic rest frame from the laboratory frame.}
\label{fig:angle}
\end{figure}

There are two analyses~\cite{colin,belle_rho} which use single tau decays
into $K\pi\nu_{\tau}$ and $\pi\pi^0$ final state using the technique
proposed by J.H.~K\"{u}hn and E.~Mirkes~\cite{kuhn,tsai}. It can be 
shown that for the decays of unpolarized tau leptons the CP-odd part of the
probability density is proportional to:
\be
P_{odd} \sim \Im(\Lambda) \Im(e^{i\delta}) \cos\beta \cos\psi,
\ee
where $\Im(\Lambda)$ is an imaginary part of the overall Higgs coupling 
(see Eq.~\ref{eq:Higgs_lambda}), $\delta$ is a relative strong phase between
$W$ and Higgs exchange diagrams. $\beta$ and $\psi$ are 
angles between a hadron, $\tau$ lepton and direction of the hadronic
rest frame in the laboratory frame (see Fig.~\ref{fig:angle}). 
Therefore, the distributions of $\cos\beta \cos\psi$ for $\tau^+$ 
and $\tau^-$ decays are given by:
\be 
\frac{dN(\tau^+)}{d\cos\beta \cos\psi} = const + 
              c_1 \Im(\Lambda) \cos\beta \cos\psi,
\label{eq:dnplus}
\ee
and
\be 
\frac{dN(\tau^-)}{d\cos\beta \cos\psi} = const + 
              c_1 \Im(\Lambda^*) \cos\beta \cos\psi,
\label{eq:dnminus}
\ee
where $const$ is contribution from the standard, CP-even part of the
probability density and $c_1$ is a proportionality constant. Asymmetry, 
defined as a difference between Eqs.~\ref{eq:dnplus} and \ref{eq:dnminus} 
normalized by the sum of these distributions is:
\be
A_{cp} = \frac{c_1}{const} \Im(\Lambda) \cos\beta\cos\psi.
\label{eq:cosa_obs}
\ee
Non-zero slope in $\cos\beta\cos\psi$ distribution indicates CP violation.
The value of $\Im(\Lambda)$ can be determined if empirical 
coefficient $c_1/const$ is known.\footnote{Can be estimated in Monte Carlo 
simulation.}
%%%%%%%%%%%%%%%%%%%%%%%%%%%%%%%%%%%%%%%%%%%%%%%%%%%%%%%%%%%%%%%%%% 
\subsection{CLEO $\tau\to K\pi\nu_{\tau}$ results}
\begin{figure}
\center
\psfig{figure=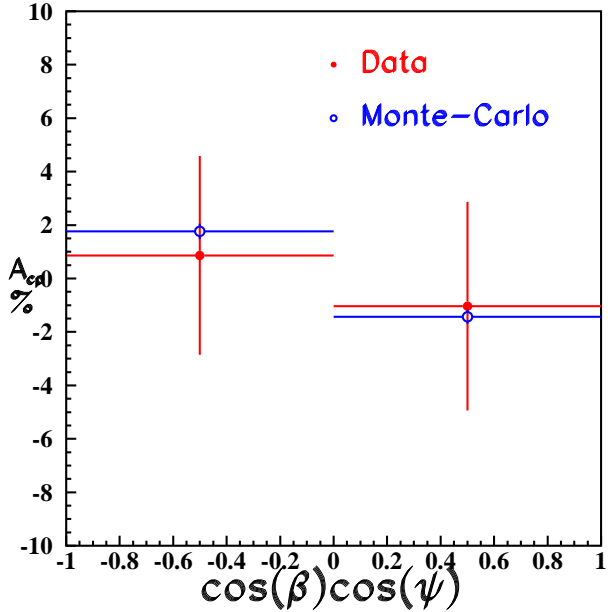,height=3.5in}
\caption{Asymmetry $A_{cp}$ as a function of $\cos\beta\cos\psi$ (CLEO).}
\label{fig:cleo_single}
\end{figure}
The first search for CP violation in tau decays is performed
in the $\tau\to K\pi\nu_{\tau}$ mode~\cite{colin}.
Data collected from $e^+e^-$ collisions at a center of 
mass energy ($\sqrt{s}$) of 10.6 GeV with the CLEO II detector
\cite{CLEOII} at the Cornell Electron Storage Ring (CESR) is used in this
analysis. 
The total integrated luminosity of the data sample is 4.8 fb$^{-1}$, 
corresponding to the production of $4.4\times10^6$ tau pairs. 
A $\tau^- \to K_s^0 h^- \nu_{\tau}$, $K_s^0\to\pi^-\pi^+$
event sample is selected since three charged tracks in the final state
are well measured. Here $h^-$ is a charged pion or kaon. The 
standard selection criteria is used to reconstruct 
$\tau\to K_s^0 h \nu_{\tau}$ by CLEO in previous analysis~\cite{Kshort}. 
\par
Observed asymmetry $A_{cp}$ (see Eq.~\ref{eq:cosa_obs}) in the data sample
for the positive and for the negative values of $\cos\beta\cos\psi$ after 
subtraction of the background is given in Table~\ref{tab:cleo_result}.
The distribution of $A_{cp}$ for the data and Standard Model Monte Carlo
is illustrated in Fig.~\ref{fig:cleo_single}.
\begin{table}
\begin{center}
\caption{Asymmetry $A_{cp}$ for negative and positive values of 
$\cos\beta\cos\psi$, CLEO Collaboration.}\label{tab:cleo_result}
\vspace{0.2cm}
\begin{tabular}{|c|r|} \hline
$\cos\beta\cos\psi$ & $A_{cp}$ \\ \hline
$\cos\beta\cos\psi< 0$ &  $0.009\pm0.038$ \\
$\cos\beta\cos\psi> 0$ &  $-0.010\pm0.039$ \\ \hline
\end{tabular}
\end{center}
\end{table}
%\vspace*{3pt}
\par
To relate the observed asymmetry $A_{cp}$ to the imaginary part of 
the coupling constant $\Lambda$, the value of the proportionality 
coefficient $c_1$ (see Eq.~\ref{eq:cosa_obs}) is estimated using 
modified TAUOLA package~\cite{KORALB}. No CP violation is observed and
the following upper limit on CP-violation is set:
\be
|\Im(\Lambda)| < 1.7, \mbox{ at 90\% C.L.}
\label{eq:cleo_result}
\ee
%%%%%%%%%%%%%%%%%%%%%%%%%%%%%%%%%%%%%%%%%%%%%%%%%%%%%%%%%%%%%
\subsection{BELLE $\tau\to \pi\pi^0\nu_{\tau}$ results}
There is an ongoing search~\cite{belle_rho} by the BELLE Collaboration
for CP violation 
in $\tau\to\pi\pi^0\nu_{\tau}$ mode using the same technique. 
The data sample used in this analysis has been collected in 
$e^-e^+$ collisions at a center of mass energy of 10.6 GeV with 
the Belle detector at the KEK asymmetric energy collider (KEKB)~\cite{KEKB}.
\begin{figure}
\center
\psfig{figure=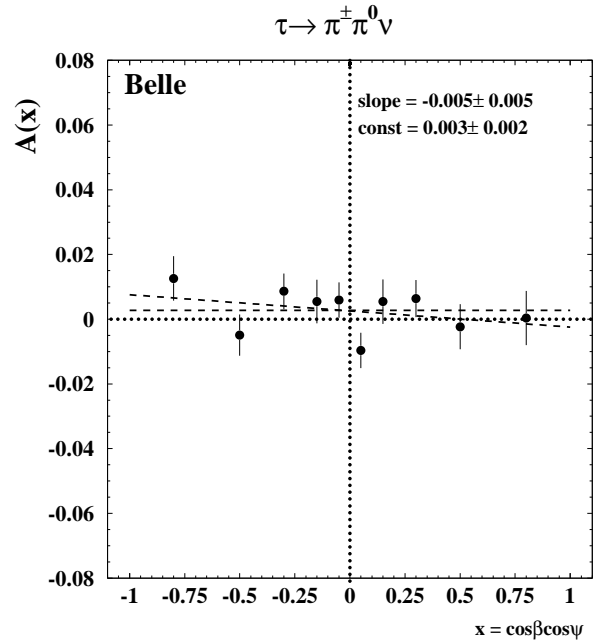,height=3.5in}
\caption{Asymmetry $A_{cp}$ as a function of $\cos\beta\cos\psi$.
The dashed lines are fits with a straight line or a constant(Belle).}
\label{fig:belle_single}
\end{figure}
The total integrated luminosity accumulated during this period is 
6.7 fb$^{-1}$, corresponding to 6.2 million tau pairs. The final sample of 
$\tau \to \pi\pi^0\nu_{\tau}$ contains $2.6\times10^5$ events. 
Using this sample the asymmetry from $\tau^+$
and $\tau^-$ decays has been measured in two intervals of $\cos\beta\cos\psi$
similarly to the previous section. The values of the observed asymmetry are
given in Table~\ref{tab:belle_result}.
\begin{table}
\begin{center}
\caption{Asymmetry $A_{cp}$ for negative and positive values of 
$\cos\beta\cos\psi$, Belle collaboration.}\label{tab:belle_result}
\vspace{0.2cm}
\begin{tabular}{|c|r|} \hline
$\cos\beta\cos\psi$ & $A_{cp}$ \\ \hline
$\cos\beta\cos\psi< 0$ &  $ 0.0056\pm0.0027$ \\
$\cos\beta\cos\psi> 0$ &  $-0.0004\pm0.0027$ \\ \hline
\end{tabular}
\end{center}
\end{table}
\noindent
The distribution of the asymmetry is illustrated in 
Fig.~\ref{fig:belle_single}. The value of the asymmetry is consistent
with previous CLEO result and indicates no CP violation in 
$\tau\to\pi\pi^0\nu_{\tau}$ decays within measurement errors:
\be
|A_{cp}| < 0.010, \mbox{ at 90 \% C.L..}
\ee 
The limit on the imaginary part of the scalar coupling will be set in
the near future.
%%%%%%%%%%%%%%%%%%%%%%%%%%%%%%%%%%%%%%%%%%%%%%%%%%%%%%%%%%%%%%%%%%%
\subsection{Summary of searches in single tau decays}
The first search for CP violation in $\tau$ decays is done in 
$\tau\to K\pi\nu_{\tau}$ mode and provides for the first time a restriction on
the imaginary part of a scalar coupling in tau decays:
\be
|\Im(\Lambda)| < 1.7, \mbox{ at 90\% C.L..}
\ee
Analysis using a single tau decaying into $\pi\pi^0\nu_{\tau}$ final state
is in progress and current upper limit on the CP-violating asymmetry is:
\be
|A_{cp}| < 0.010, \mbox{ at 90 \% C.L..}
\ee
This limit will be improved with better understanding of the systematics
and an increase of the data sample. A conversion of the above upper limit
to the imaginary part of the scalar coupling will be done in the near future.
\par
Searches for CP violation in a single $\tau$ decay are sensitive only
to a spin-independent CP-violating term in the decay rate. Other
spin-dependent terms are proportional to tau polarization vector
and, therefore, average to zero. However, the 
situation is different for decays of tau pairs produced 
in $e^+e^-$ annihilations, where the parent virtual photon introduces 
correlations of the $\tau^+$ and $\tau^-$ spins. This correlation
provides an additional constraint which permits to be sensitive
to all CP-odd terms of the decay rate and maximizes the sensitivity
to possible CP-violating effects.
In the following, I discuss an analysis~\cite{CLEO_recent} without 
these disadvantages.
%
%%%%%%%%%%%%%%%%%%%%%%%%%%%%%%%%%%%%%%%%%%%%%%%%%%%%%%%%%%%%%%%%%%%
\section{CP violation in correlated tau decays}
\begin{figure}
\center
\psfig{figure=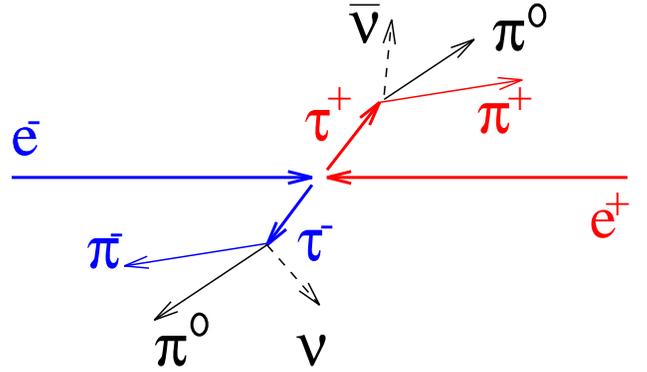,height=2in}
\caption{Correlated decay of tau pair into $\pi\pi^0\nu_{\tau}$ each.}
\label{fig:corr_decay}
\end{figure}
\begin{figure}
\center
\psfig{figure=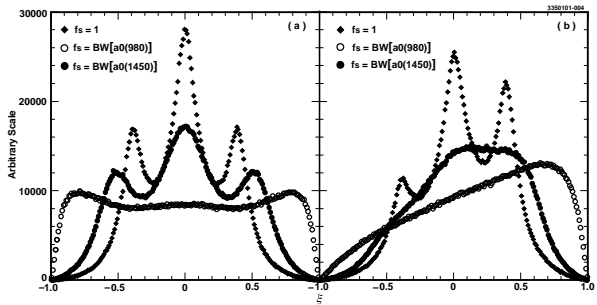,height=1.6in}
\caption{Optimal observable $\xi$ for (a) Monte Carlo with no 
CP violation and (b) Monte Carlo with a maximal CP violation
$\Im(\Lambda)=1$.}
\label{fig:fs}
\end{figure}

To maximize the sensitivity of our search an optimal 
variable is constructed, first proposed~\cite{optim1} by D.~Atwood and 
A.~Soni, $\xi$, with the smallest associated statistical error.
The variable is equal to the ratio of the CP-odd and CP-even parts 
of the total cross section assuming that the absolute value of the coupling 
$\Lambda$ is unity:
\be
\label{eq:xi_opt}
	\xi = \frac{P_{odd}}{P_{even}},
\ee 
where $P_{odd}$ and $P_{even}$ are the CP-odd and CP-even parts
of the total cross section. These terms are functions of the vector
and scalar hadronic currents. A vector hadronic current is parameterized 
by the relative momentum between the charged and neutral pions 
multiplied by the vector form factor, $f_v$, described by 
$\rho$ Breit-Wigner shape:
\begin{equation}
\label{eq:BW}	
 f_v = \frac{-m^2}{s - m^2 + i m \Gamma(s)},
\end{equation}
where $s$ is a squared invariant mass of two pions, $m_{\pi}$ is a pion mass,
and $m$ and $\Gamma(s)$ are the mass and the momentum-dependent width of 
the resonance, respectively. The latter is defined as:
\begin{equation}
\Gamma(s) = \left \{ \begin{array}{ll}
\frac{m}{\sqrt{s}}(\frac{s-4m_{\pi}}{m^2-4m_{\pi}^2})^{3/2}
  & \mbox{if $s>(2m_{\pi})^2$} \\
0 & \mbox{elsewhere.}
\end{array}
\label{eq:BW_width}
\right .
\end{equation}
Here, the contribution from the $\rho^\prime$ 
resonance~\cite{rhoprime} is neglected.
The scalar hadronic current is parameterized as a product of a 
dimensional quantity, $M=1$ GeV/$c^2$ providing overall normalization, 
and a scalar form factor $f_s$. The choice of the scalar form factor is not 
unambiguous. Three possible cases are studied: one with $f_s = 1$, 
the second with $f_s$ described by the $a_0(980)$ Breit-Wigner shape, 
and the third with $f_s$ described by $a_0(1450)$ Breit-Wigner shape 
(see Eq.~\ref{eq:BW}) with a width given by:
\begin{equation}
\Gamma(s) = \left \{ \begin{array}{ll}
\frac{m}{\sqrt{s}}(\frac{s-4m_{\pi}}{m^2-4m_{\pi}^2})^{1/2}
  & \mbox{if $s>(2m_{\pi})^2$} \\
0 & \mbox{elsewhere.}
\end{array}
\right .
\end{equation} 
Since $P_{odd}$ is proportional to the imaginary part of $\Lambda$,
then it can be expressed as: 
\begin{equation}
 P_{odd}(\Lambda) = \Im(\Lambda) P_{odd}(1),
\end{equation}
and
\begin{equation}
\label{eq:xi_ave}
 <\xi> = \Im(\Lambda) \int \frac{P_{odd}(1)^2}{P_{even}} dLips. 
\end{equation}
The integral in Eq.~\ref{eq:xi_ave} is always larger than or equal to
zero, and equality occurs only if the odd part of the cross section 
vanishes. Therefore, the average value of $\xi$ is  
proportional to the imaginary part of the Higgs coupling constant and 
is positive if $\Im(\Lambda) > 0$ and negative if $\Im(\Lambda) < 0$. 
Monte Carlo simulation of the $\xi$ distributions for the three choices 
of the scalar form factors and no CP violation are shown in 
Fig.~\ref{fig:fs}(a). The same distributions for the CP violating 
case $\Im(\Lambda)=1$ are shown in Fig.~\ref{fig:fs}(b).
The structure in these distributions is due to the resonant structure in the 
vector and scalar form factors.
\par
To relate the observed mean value of the optimal observable $<\xi>$ to
the imaginary part of the coupling constant $\Lambda$, the $\Im(\Lambda)$
dependence of $<\xi>$ must be known. To the first order 
the mean value is proportional to $\Im(\Lambda)$ with a 
proportionality coefficient $c$:
\be
 <\xi> = c \times \Im(\Lambda).
\label{eq:xi_powers}
\ee
The value of this coefficient is estimated using Monte Carlo 
simulation as described in the Section~\ref{sec:calibration}.
%%%%%%%%%%%%%%%%%%%%%%%%%%%%%%%%%%%%%%%%%%%%%%%%%%%%%%%%%%%%%%%%%
\subsection{Data and Monte Carlo Samples}
The data used in this analysis were collected at the Cornell 
Electron Storage Ring (CESR) at or near the energy of the $\Upsilon(4S)$. 
The data correspond to a total integrated luminosity of 13.3 fb$^{-1}$ and 
contain 12.2 million $\tau^+\tau^-$ pairs. Versions of the CLEO detector 
employed here are described in Refs.~\cite{CLEOII,CLEOII5}.
From this data sample, events consistent with $e^+e^-\to\tau^+\tau^-$ 
interactions are selected where each 
$\tau$ decays into the $\pi\pi^0\nu_\tau$ final state.
The event selection follows mostly the procedure developed originally 
for the study of the $\tau\to\rho\nu_{\tau}$ decays~\cite{rho-rho}.
In the following section it is shown that the event selection criteria 
do not introduce artificial CP-violating effects.
\par
To estimate backgrounds large samples of Monte Carlo events are analyzed 
following the same procedures that are applied to the actual CLEO data.  
All non-tau backgrounds are found to be negligible. 
The main remaining background is due to the $\tau$-pair events in which 
one of the $\tau$'s decays into $\rho\nu_{\tau}$ while the other decays 
into $\pi 2\pi^0\nu_{\tau}$ and the photons from one of the $\pi^0$'s 
are not detected. 
The contamination from this background source is estimated to be 5.2\%. 
The second largest background contribution of 2.1\% is due to one of 
the $\tau$'s decaying into the $K^*\nu_\tau$ final state producing 
a neutral pion plus a charged kaon which is mistaken 
for a pion. All other tau decays provide much smaller contributions 
with the largest being less than 0.7\%. The total background 
contamination from tau decays is estimated to be 9.9\%. 
This background neither introduce a bias in the $<\xi>$ distribution
nor modifies the value of the coefficient $c$, defined
in Eq.~\ref{eq:xi_powers}.
%%%%%%%%%%%%%%%%%%%%%%%%%%%%%%%%%%%%%%%%%%%%%%%%%%%%%%%%%%%%%%%
\subsection{Calibration}\label{sec:calibration}
To relate the observed mean value of the optimal observable $<\xi>$ to
the imaginary part of the coupling constant $\Lambda$, coefficient
$c$ must be determined. It is estimated using several signal Monte 
Carlo samples generated\footnote{These events are generated with full 
GEANT-based detector simulation~\cite{GEANT} 
and pattern recognition software.} 
with different values of $\Im(\Lambda)$. For each sample we
calculate the average value of the optimal observable $\xi$ 
and plot it as a function of $\Im(\Lambda)$. For each form of the 
scalar component, the calculated asymmetry distribution 
is fit to a straight line to obtain the calibration coefficients.
To check that the selection criteria do not create an artificial 
asymmetry the mean value of the optimal observable is calculated for 
Standard Model Monte Carlo samples for each choice of the scalar 
form factor. These values along with calculated coefficients 
$c$ are listed in Table~\ref{tab:coeff}.
\begin{table}
\begin{center}
\caption{Average values of the optimal observable $<\xi>$ for the 
Standard Model Monte Carlo and the proportionality coefficient $c$ 
for the CP asymmetry fits for different scalar form factors.}\label{tab:coeff}
\vspace{0.2cm}
\begin{tabular}{|c|r|r|} \hline
Form factor, $f_s$       & $<\xi>$, $10^{-3}$     & $c$, $10^{-3}$ \\ \hline
1                        & $ 0.7\pm0.6$           & $66.8\pm4.3$   \\ 
$BW(a_0(980))$           & $ 1.0\pm1.1$           & $586.4\pm19.4$ \\ 
$BW(a_0(1450))$          & $ 0.5\pm0.8$           & $145.8\pm7.3$  \\ \hline
\end{tabular}
\end{center}
\end{table}
\noindent
For all three form factors, the mean value of $\xi$ for the 
Standard Model Monte Carlo sample is consistent with zero within 
its statistical error. Therefore, event selection criteria do not 
introduce an artificial asymmetry in the $\xi$ distribution.
The coefficients from Table~\ref{tab:coeff} are used to 
calculate $\Im(\Lambda)$. 
%%%%%%%%%%%%%%%%%%%%%%%%%%%%%%%%%%%%%%%%%%%%%%%%%%%%%%%%%%%%%%%%%
\subsection{Observed mean values}
\label{ss_results}
\begin{figure}
\center
\psfig{figure=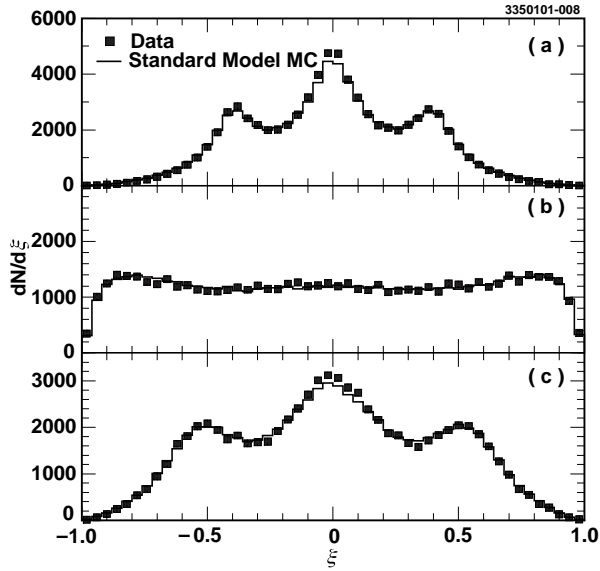,height=3in}
\caption{The distribution of 
the CP violation sensitive variable $\xi$ for the data (dots) compared to 
the Standard Model Monte Carlo prediction (solid line) for (a) $f_s$=1, 
(b) $f_s=BW(a_0(980))$ and (c) $f_s=BW(a_0(1450))$.}
\label{fig:data}
\end{figure}

For each choice of the scalar form factor, a distribution
in $\xi$ is obtained. These distributions are shown in Fig.~\ref{fig:data}, 
with those from the Standard Model Monte Carlo simulations overlaid. 
From these distributions the mean values $<\xi>$ are computed after 
subtracting the average value for Standard Model Monte Carlo, 
which are reported in the first column of Table~\ref{tab:results}. 
In each case, the appropriate empirically-determined coefficient
given in Table~\ref{tab:coeff} is used to derive a value for the imaginary 
part of the Higgs coupling $\Lambda$, as described in the preceding 
section. These values along with the $90\%$ confidence limits on 
$\Im(\Lambda)$ are reported in the second and third columns of 
Table~\ref{tab:results}.
\begin{table}
\begin{center}
\caption{Average value of the optimal observable $\xi$ after subtracting the 
average value for Standard Model Monte Carlo, calculated value of 
$\Im(\Lambda)$ and 90\% C.L. intervals on $\Im(\Lambda)$.}\label{tab:results}
\vspace{0.2cm}
\begin{tabular}{|c|r|r|r|} \hline
$f_s$                 & $<\xi>$, $10^{-3}$ & $\Im(\Lambda)$, $10^{-2}$ 
                      & $\Im(\Lambda)$, 90\% C.L. \\ \hline
$1$                   & $-0.8\pm1.4$ & $-1.2\pm2.1$ & $(-0.046; 0.022)$ \\
$a_0(980)$        & $-0.6\pm2.4$ & $-0.1\pm0.4$ & $(-0.008; 0.006)$ \\
$a_0(1450)$       & $ 0.2\pm1.7$ & $ 0.1\pm1.2$ & $(-0.019; 0.021)$ \\ 
\hline
\end{tabular}
\end{center}
\end{table}
%%%%%%%%%%%%%%%%%%%%%%%%%%%%%%%%%%%%%%%%%%%%%%%%%%%%%%%%%%%%%%%%%
\subsection{Summary}
\label{ss_summary}
No sizable systematic errors are found\footnote{The detailed description
of the most significant effects is given elsewhere~\cite{CLEO_recent}.} 
which could alter the limits shown in Table~\ref{tab:results}. 
Within our experimental precision no significant 
asymmetry of the optimal variable is observed and, therefore, 
no CP violation in $\tau$ decays. Due to the uncertainty in the 
choice of the scalar form factor the most conservative 90\% confidence 
limits are used corresponding to $f_s=1$:
$$ -0.046 <\Im(\Lambda) < 0.022, \mbox{ at 90\% C.L.}. $$
These limits include the effects of possible systematic errors.
%%%%%%%%%%%%%%%%%%%%%%%%%%%%%%%%%%%%%%%%%%%%%%%%%%%%%%%%%%%%%%%%%
\subsection{Pseudo-helicity method}
In this section I discuss another method to search for 
CP violation in $\tau$ decays. It is more intuitive but less
sensitive than the optimal observable method described above.
\par
The helicity angle, $\theta_{\pi\pi}$, is defined as  
the angle between the direction of the charged pion in the $\pi\pi^0$ 
rest frame and the direction of the $\pi\pi^0$ system in the
$\tau$ rest frame. In Standard Model, the helicity angle is expected 
to have a distribution corresponding to a vector exchange:
\begin{equation}
 \frac{dN}{d~cos\theta_{\pi\pi}} \sim a + b\cos^2\theta_{\pi\pi}.
 \label{eq:helDIS}
\end{equation}
For scalar-mediated decays, there is an additional term proportional 
to $\cos\theta_{\pi\pi}$ that corresponds to the S-P wave interference 
and linearly proportional to the scalar coupling constant $\Lambda$.
In general, $\Lambda$ is complex and the term linear in $\cos\theta_{\pi\pi}$
is proportional to the real and imaginary parts of the scalar coupling 
with coefficients $c_1$ and $c_2$, respectively:
\begin{equation}
 \frac{dN}{d~cos\theta_{\pi\pi}} \sim a + 
                     c_1 \Re(\Lambda) \cos\theta_{\pi\pi}
                   + c_2 \Im(\Lambda) \cos\theta_{\pi\pi}
                                     + b\cos^2\theta_{\pi\pi}.
 \label{eq:helDIS2}
\end{equation}
The observation of the terms proportional to cosine of the helicity
angle would indicate the scalar exchange in the tau decays. 
\par
In order to calculate the helicity angle tau rest frame must be known.
Due to the unobserved neutrino, the tau rest frame can only be reconstructed
with a two-fold ambiguity. Such ambiguity can be avoided by using the 
pseudo-helicity angle, $\theta^*$. This pseudo-helicity angle is obtained
by replacing the tau rest frame with the laboratory rest frame where it 
is defined as an angle between the direction of $\pi^\pm$ in the $\pi\pi^0$ 
rest frame and the direction of the $\pi\pi^0$ system in the lab frame.
The difference between the pseudo-helicity distributions for the
$\tau^+$ and $\tau^-$ decays is expected to have the same form as 
given by Eq.~\ref{eq:helDIS2} but with a different numerical coefficients.
\par
The term including $\Im(\Lambda)$ changes sign for
tau leptons of opposite charges. Therefore, the difference of the 
pseudo-helicity distributions for positive and negative tau leptons 
has the term linear in $\cos\theta^*$ proportional to the imaginary part 
of the scalar coupling $\Lambda$ only:
\begin{equation}
 \frac{dN(\tau^-)}{d~cos\theta^*} -  \frac{dN(\tau^+)}{d~cos\theta^*}
 \sim 2 c_2 \Im(\Lambda) \cos\theta^*.
 \label{eq:hel_observable2}
\end{equation}
The presence of this term indicates CP violation.
\par
\begin{figure}
\center
\psfig{figure=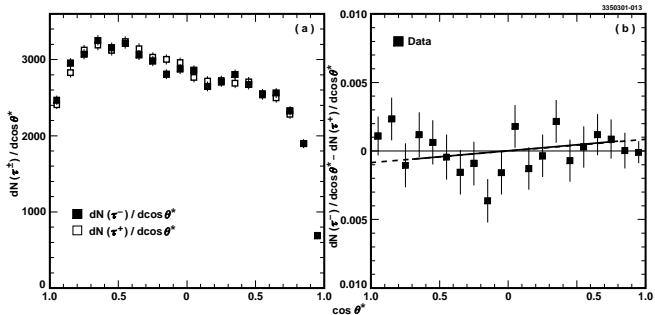,height=1.65in}
\caption{(a) Pseudo-helicity
distribution for $\tau^-$ and $\tau^+$ in data, (b) difference
between pseudo-helicity distributions for $\tau^-$ and $\tau^+$ in data.
Solid line is a linear fit and dashed lines show extrapolation to the
region excluded from the fit.}
\label{fig:hel_data}
\end{figure}

In this study, the same data sample is used as for the optimal observable
analysis with the same selection criteria. The pseudo-helicity distribution
for $\tau^-$ and $\tau^+$ is given in Fig.~\ref{fig:hel_data}(a).
\par
The structure in Fig.~\ref{fig:hel_data}(a) is due to the variation in the 
efficiency as a function of charged pion momentum and $\pi^0$ energy.
To obtain the product of the imaginary part of the scalar 
coupling $\Lambda$ and a linearity coefficient $c_2$ 
(see Eq.~\ref{eq:hel_observable2}), the difference 
of the two pseudo-helicity distributions 
for negative and positive tau leptons is fit to a first order polynomial. 
To minimize systematic effects due to soft
pion reconstruction the fit is performed in the region of 
$-0.7<\cos\theta^*<0.8$, which
corresponds to pions with momentum higher than 0.3 GeV/$c$.
The obtained value of the slope for the data distribution is:
$$ c_2 \Im(\Lambda) = (4.2 \pm 3.6) \times 10^{-4}$$
and illustrated in Fig.~\ref{fig:hel_data}(b).
It is consistent with zero within statistical error.
\par
To determine the proportionality coefficient $c_2$ the procedure 
described in Section~\ref{sec:calibration} is followed. 
The slope of the difference of pseudo-helicity distributions
is calculated after applying selection criteria. Five samples
of 200,000 signal Monte Carlo events generated with different values
of $\Im(\Lambda)$ are used. Then, the calculated slope dependence
as a function of $\Im(\Lambda)$ is fit to a straight line to obtain $c_2$:
\be
 c_2 = (107.5 \pm 12.6)\times 10^{-4}.
\ee
This coefficient is used to obtain the value of the imaginary part of the
scalar coupling $\Im(\Lambda)$ after the correction of the slope by
Monte Carlo:
\be
\Im(\Lambda) = 0.028 \pm 0.037, 
\ee
and
\be
-0.033 < \Im(\Lambda) < 0.089 \mbox{ at 90\% C.L.}
\label{eq:hel_result_im}
\ee
As expected, the limit on the $\Im(\Lambda)$ is less strict than 
the one obtained using the optimal observable.
%%%%%%%%%%%%%%%%%%%%%%%%%%%%%%%%%%%%%%%%%%%%%%%%%%%%%%%%%%%%%%%%%%%
\section{Summary}
I have discussed three searches for CP violation
in the decays of tau leptons. No evidence for CP violation is
observed within the experimental precision. First two 
analyses search for non-zero CP-violating asymmetry in the single 
tau decays to the $K\pi\nu_{\tau}$ and 
$\pi\pi^0\nu_{\tau}$ final states. The following 
upper limit on the imaginary part of the MHDM scalar coupling 
constant $\Lambda$ is set in $\tau\to K\pi\nu_{\tau}$ decays:
\be 
|\Im(\Lambda)| < 1.7 \mbox{ at 90\% C.L.},
\label{eq:colin}
\ee
The value of the asymmetry for $\tau\to\pi\pi^0\nu_{\tau}$ is:
\be
|A_{cp}| < 0.010, \mbox{ at 90 \% C.L.}
\ee 
The conversion of this asymmetry to the value of $\Im(\Lambda)$
will be done in the near future. 
\par
The third analysis used optimal variable to search for CP violation 
in correlated tau decays decaying into $\pi\pi^0\nu_{\tau}$ final state each. 
The following confidence limits on $\Im(\Lambda)$ are set: 
\be
-0.046 < \Im(\Lambda) < 0.022 \mbox{ at 90\% C.L.}
\label{eq:maravin}
\ee 
All these results agree with each other and the limits on imaginary part
of the coupling constant $\Lambda$ restrict the size of the 
contribution of multi-Higgs-doublet model diagrams to the $\tau$ lepton decay.
%\section*{Acknowledgments}

\end{document}